\documentclass[sigconf]{acmart}

\AtBeginDocument{%
  \providecommand\BibTeX{{%
    \normalfont B\kern-0.5em{\scshape i\kern-0.25em b}\kern-0.8em\TeX}}}

\setcopyright{rightsretained}
\copyrightyear{2022}
\acmYear{2022}
\acmConference{SIGGRAPH '22 Emerging Technologies }{August 07-11, 2022}{Vancouver, BC, Canada}\acmBooktitle{Special Interest Group on Computer Graphics and Interactive Techniques Conference Emerging Technologies (SIGGRAPH '22 Emerging Technologies ), August 07-11, 2022}\acmDOI{10.1145/3532721.3535568}
\acmISBN{978-1-4503-9363-8/22/08}

\begin{document}

\title{Sense of Embodiment Inducement for People with Reduced Lower-body Mobility and Sensations with Partial-Visuomotor Stimulation}

\author{Hyuckjin Jang}
\authornote{Both authors contributed equally to this research.}
\email{jang5s@kaist.ac.kr}
\orcid{0000-0003-2398-4922}
\affiliation{%
  \institution{KAIST\country{Republic of Korea}}
}

\author{Taehei Kim}
\email{hayleyy321@kaist.ac.kr}
\authornotemark[1]
\affiliation{
  \institution{KAIST\country{Republic of Korea}}
}

\author{Seo Young Oh}
\email{seoyoung.oh@kaist.ac.kr}
\affiliation{
  \institution{KAIST\country{Republic of Korea}}
}

\author{Jeongmi Lee}
\email{jeongmi@kaist.ac.kr}
\affiliation{
  \institution{KAIST\country{Republic of Korea}}
}

\author{Sunghee Lee}
\email{sunghee.lee@kaist.ac.kr}
\affiliation{
  \institution{KAIST\country{Republic of Korea}}
}
\author{Sang Ho Yoon}
\email{sangho@kaist.ac.kr}
\affiliation{
  \institution{KAIST\country{Republic of Korea}}
}

\renewcommand{\shortauthors}{Jang and Kim, et al.}

\begin{abstract}
To induce the Sense of Embodiment~(SoE) on the virtual 3D avatar during a Virtual Reality~(VR) walking scenario, VR interfaces have employed the visuotactile or visuomotor approaches. However, people with reduced lower-body mobility and sensation~(PRLMS) who are incapable of feeling or moving their legs would find this task extremely challenging. Here, we propose an upper-body motion tracking-based partial-visuomotor technique to induce SoE and positive feedback for PRLMS patients. We design partial-visuomotor stimulation consisting of two distinctive inputs~(\textit{Button Control} \& \textit{Upper Motion tracking}) and outputs~(\textit{wheelchair motion} \& \textit{Gait Motion}). The preliminary user study was conducted to explore subjective preference with qualitative feedback. From the qualitative study result, we observed the positive response on the partial-visuomotor regarding SoE in the asynchronous VR experience for PRLMS.
\end{abstract}

\maketitle

\section{Introduction}

With the advancement of virtual reality~(VR) and full-body motion tracking, a full-body avatar has been deployed in recent research and industrial VR interfaces. A Sense of Embodiment~(SoE), which refers to a subjective feeling of experiencing and owning a body~\cite{kilteni2012sense} has become a crucial component for the VR experience where the avatar emerges as a primary interacting medium within the VR. However, people with reduced lower-body mobility and sensation~(PRLMS) due to the external damage and subsequent loss of motor function could not experience SoE during VR walking scenarios. The main reason is that existing methods require multi-modal synchronous inputs~\cite{kokkinara2014measuring} or arm swing in a stand-up position (\cite{Cannavo2021, 10.1145/2804408.2804416, 10.1145/3152832.3152864, 10.1145/3013971.3014010}) which are not applicable to PRLMS. Previous studies mainly attempted to use passive viewing of the VR environment, the visual-tactile method, or treadmills to induce the SoE of PRLMS. However, a novel barrier-free approach is needed to encompass the broad users, including PRLMS. This study aims to find an effective and economical method that can further increase the SoE only with controllers. 

\begin{figure}[hb!]
    \centering
    \vspace{-0.15cm}
    \includegraphics[width=\columnwidth]{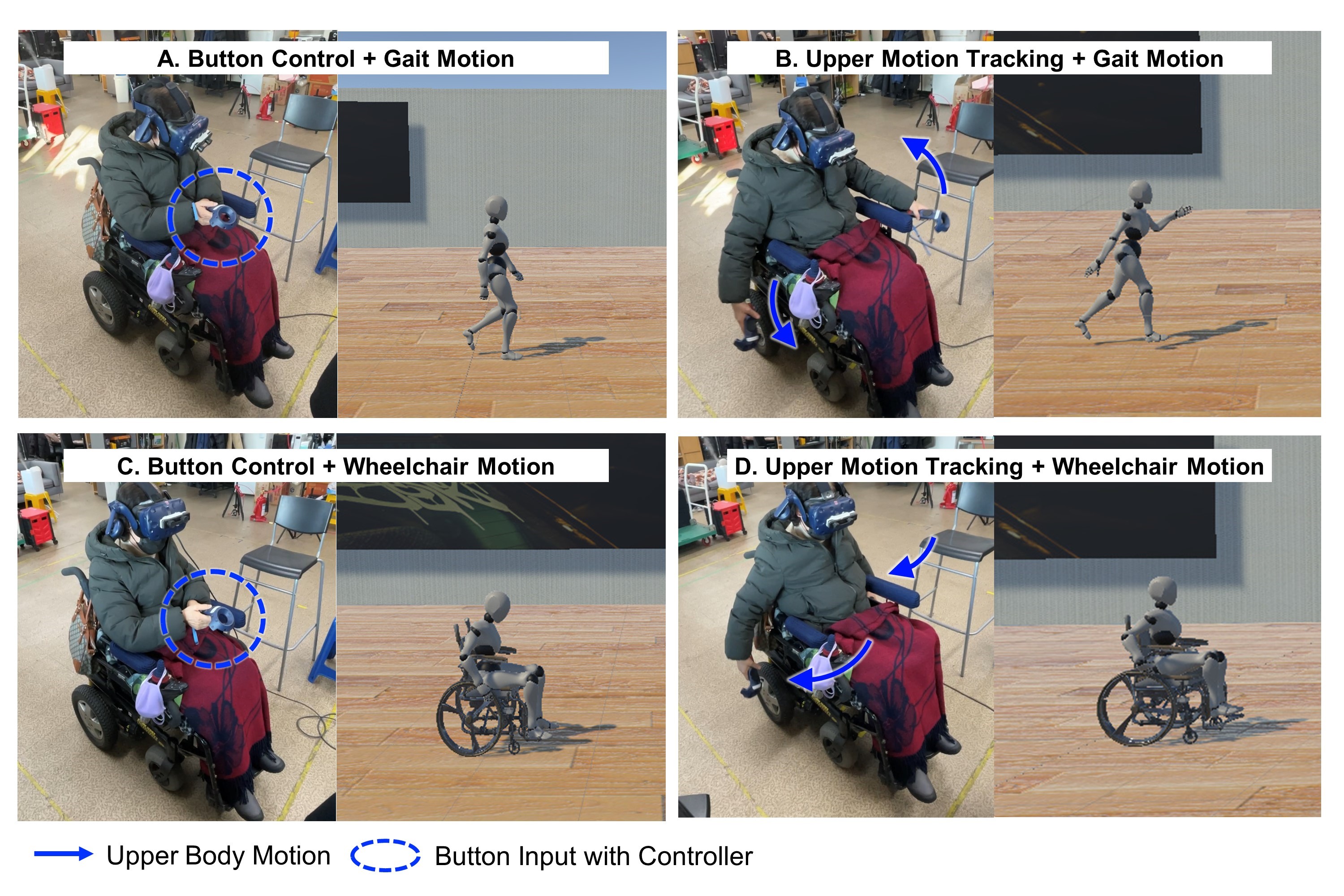}
    \vspace*{-7mm}
    \caption{Participant with lower limb paralysis carries out a user study using proposed system where corresponding third person perspective avatars are shown together.}
    \label{Fig:experiment}
    \vspace{-0.15cm}
\end{figure}

In this work, we propose \textit{partial-visuomotor stimulation} which automatically generates lower body motion animation from upper motion tracking. The upper motion tracking is based on headset and hand controller movement tracking. We seek to find a way to enhance SoE inducement in situations where PRLMS have inevitably different body representations compared to full-body walking avatars in VR~(Figure~\ref{Fig:experiment}).

\section{Implementation and Apparatus}
\begin{figure}[htb!]
    \centering
    \vspace{-0.15cm}
    \includegraphics[width=\columnwidth]{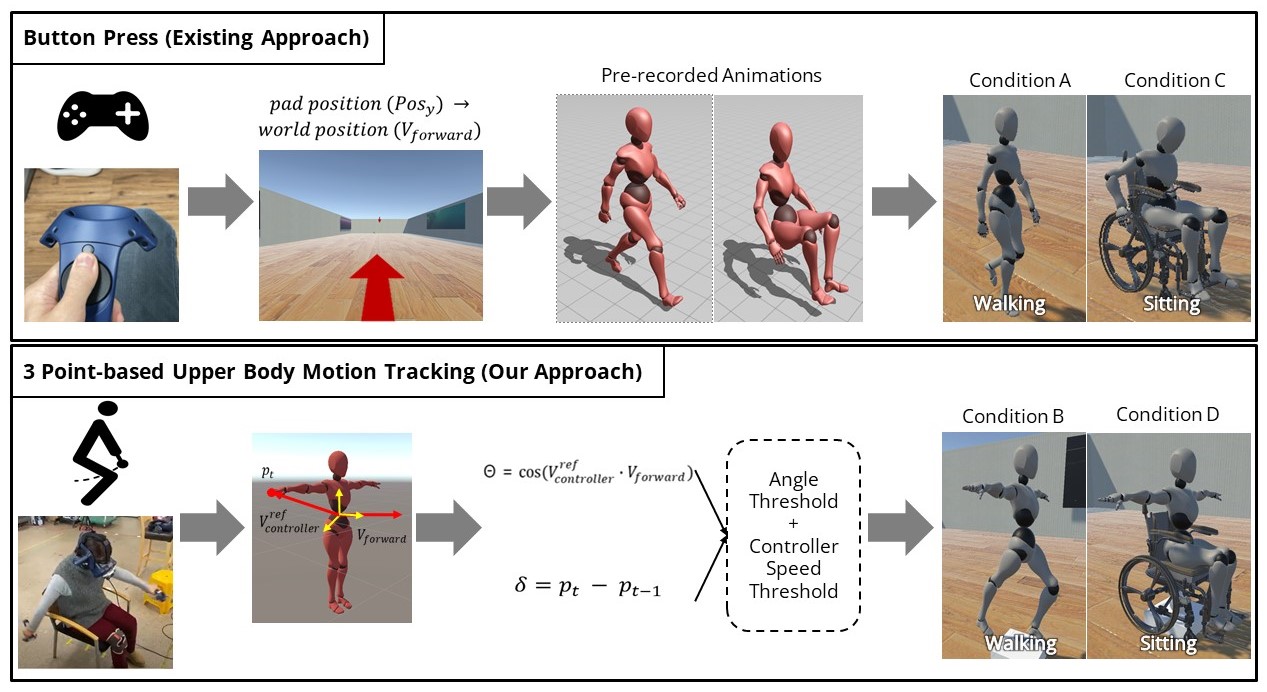}
    \vspace*{-6mm}
    \caption{System Overview. We propose a 3 point-based upper body motion tracking for SoE inducement for People with Reduced Lower-body Mobility and Sensations.}
    \label{fig:systemoverview}
    \vspace{-0.15cm}
\end{figure}

Figure~\ref{fig:systemoverview} illustrates the implementation of four different scenarios with various input (\textit{controller manipulation} \& \textit{upper body motion tracking with controllers}) and output methods (\textit{lower body motion generation on a wheelchair} \& \textit{lower body motion generation for gait}).

\begin{enumerate}
    \item \textbf{Condition A: Button Control + Gait Motion} \textit{HTC Vive} controller pad was used for the avatar manipulation where the avatar movement was limited to forward/backward directions. We adopted the automated animation from \textit{Mixamo} for the gait motion. 
    
    \item \textbf{Condition B: Upper Motion Tracking + Gait motion} For upper body motion tracking, we used SteamVR Plugin to track the movement of the headset and two controllers along with \textit{VR Final IK} Unity asset. To calculate the relative position of the hand controllers, we first define an artificial reference coordinate that hand controllers can refer to. Then, we compute the angle~($\theta$) between the hand controller location relative to the reference coordinate~($V^{ref}_{controller}$) and a forward vector~($V_{forward}$) along with the x-axis~(right vector). 
    \begin{equation}
    \label{eqn:controllerAngle}
    \theta = \cos^{-1}(V^{ref}_{controller} \cdot V_{forward})
    \end{equation}
    For the lower body gait motion, we changed the local joint angles of each leg with a time interval. Both legs change angle when the hand controller angle falls within a pre-defined range~(-130\textdegree$\sim$-90\textdegree). If the user stops moving their hands, the character also stops moving.

    \item \textbf{Condition C: Button Control + Wheelchair Motion} The \textbf{Condition C} has same input method as \textbf{Condition A}. For the wheelchair motion, we embedded the ``Wheelchair'' animation. We added ``wheel pulling hand gestures'' and ``wheel rotation'' animations when moving forward.

    \item \textbf{Condition D: Upper Motion Tracking + Wheelchair Motion} The position of the hand controller is used as an input. The user swings his hands back and forward simultaneously as if he is pushing the wheelchair. To calculate the location of the hand controller, we assigned a fixed reference similar to \textbf{Condition C}. If the angle between the location of the hand controller to the fixed reference along the x-axis (right vector) is between -130\textdegree$\sim$-100\textdegree, the moving forward command is triggered. We fixed the local rotation of leg joints to set the lower body animation as a general sit pose.
\end{enumerate}

\balance
\section{Result and conclusion}
We recruited a total of 8 participants~(6 female, 2 male) ranging from 36 to 64~(M=57, SD=9.20) who possess paralysis on a lower limb and are wheelchair users~(paraplegia due to spinal cord injury=3, lower body paralysis caused by polio=3, left hemiplegia due to cerebral infarction=1, leg amputee =1). All participants experience all experiment settings in a within-subject design.

Participants reported positive feedback on the \textbf{Condition B}: \textit{``I think the motion of standing up is much better~(P6)", "Button pressing felt convenient, but it didn't feel like I was moving my body~(P2)", "The upper body+walking was felt like real walking which makes me swinging my arms more enthusiastically in the subsequent trial~(P7)"}, and \textit{``I felt like I was walking while moving my feet in the upper motion tracking+gait motion~(P8)"}. Here, participants preferred walking virtually over riding a wheelchair, even if they were sitting in the wheelchair in the real world. \textbf{P7} mentioned \textit{``It is usually physically/mentally exhausting to ride a wheelchair. For example, I tend to be out of breath even if I ride an electric wheelchair"}. Furthermore, \textit{P8} stated that \textit{``The wheelchair conditions were not very interesting because it just simulates the real world."}

This work explores the prospective method of inducing the SoE in a VR walking scenario for PRLMS. Throughout the preliminary study with PRLMS, we observed positive feedback regarding SoE for upper motion tracking based on walking experience compared to other conditions. We expect that future research with a larger participant pool will provide richer insights. Moreover, the findings will serve as a foundation for the development of variety of upper body motion involved VR activities for PRLMS such as hiking, fishing, and climbing.

\begin{acks}
This work was supported by Korea Institute for Advancement of Technology(KIAT) grant funded by the Korea Government(MOTIE) (P0012746, The Competency Development Program for Industry Specialist) and KAIST grant (G04210059). Any opinions, findings, and conclusions or recommendations expressed in this material are those of the authors and do not necessarily reflect the views of the funding agency.
\end{acks} 


\balance
\bibliographystyle{ACM-Reference-Format}
\bibliography{Gen329_Final}

\end{document}